\documentclass{jps-cp}

\usepackage{siunitx}

\DeclareSIUnit{\neq}{n_{eq} }
\DeclareSIUnit{\C}{c}
\DeclareSIUnit{\protons}{protons}
\DeclareSIUnit\clight{\text{\ensuremath{c}}}

\title{Pixelated 3D sensors for tracking in radiation harsh environments}

\author{Jordi \textsc{Duarte-Campderros}$^{1}$ on behalf of the \textsc{RD50 Collaboration}}

\inst{$^{1}$Instituto de Fisica de Cantabria (IFCA (UC/CSIC)), Av. de los Castros s/n, Santander 39005, Spain}

\email{jorge.duarte.campderros@cern.ch}

\recdate{November 26, 2020}

\abst{The High Luminosity upgrade of the CERN Large Hadron Collider
will be able to reach a peak instantaneous luminosity of \SI{5e34}{\per\square\centi\metre\per\second}. 
The innermost detectors of the CMS and ATLAS experiments will have to cope with unprecedented requirements
on radiation hardness. At the end of the operation period, radiation levels are expected to reach values
above \SI{2.6e16}{\neq\square\per\centi\meter}. Sensors based on 3D pixel technology, with intrinsic radiation
tolerance, are the baseline option for the innermost layers of the vertex detectors of several HL-LHC 
experiments. This article gives an overview of the ongoing characterization of the pixelated 
3D sensor technology, their performance and their current development status for tracking on radiation
harsh environments. 
\kword{particle tracking detectors,HL-LHC upgrade,3D silicon pixel detectors, radiation-hard detector}}

\begin{document}
\maketitle

\section{Introduction}
The High Luminosity Large Hadron Collider (HL-LHC) project\cite{hl-lhc} is going to push their experiments to very 
stringent radiation environments. The innermost detectors of the largest HL-LHC experiments, ATLAS and
CMS, will have to cope with unprecedented radiation levels and with extremely high track densities. Their
upgraded vertex detectors will have to be able to operate in a hostile radiation environment, reaching hadron fluences 
up to about \SI{2.6e16}{\neq\square\per\centi\meter} before being replaced~\cite{ph2-cms,ph2-atlas}. Sensors based on 3D pixel 
technology are excellent candidates because of their inherent radiation tolerance. 

\begin{figure}[tbh]
    \centering
    \includegraphics[width=0.55\textwidth]{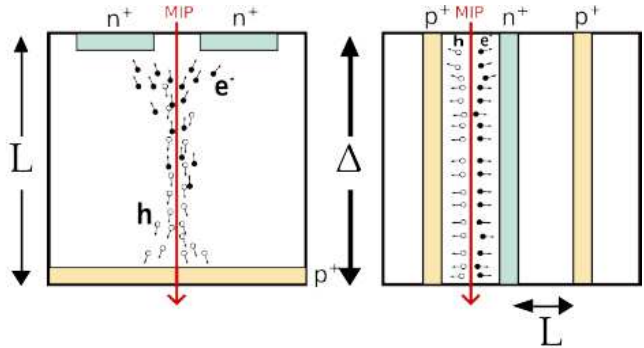}
    \caption{Schematic simplified layout of a planar (left) and a 3D (right) sensors, assuming a n-in-p layer structure. 
             Read out electrodes are bonded to the highly doped n$^+$ thin region (colored in green). Electrodes
             bonded to the ohmic layers p$^+$ (colored in yellow) are needed to polarize the sensor. The sensor
             thickness, $\Delta$, determines how many charges are created int the sensor when a particle passes 
             through, while the inter-electrode distance, $\text{L}$, defines the maximum distance a charge carrier 
             should travel before being collected. It can be observed that in a planar sensor $\text{L}=\Delta$, but 
             in a 3D sensor $\text{L} << \Delta$.}
    \label{fig:3d-planar}
\end{figure}

A 3D sensor is built by inserting highly doped columns into the bulk to form the P-N junctions, 
decoupling in that way the inter-electrode distance from the active sensor thickness 
(see Fig.~\ref{fig:3d-planar}). The thickness is defining the amount of charge generated but is in the case of 
3D sensors independent of the distance between the electrodes, respectively the depletion voltage. The inter-electrode 
distance can thus be optimized independent of the amount of charge deposited by an ionizing particle, as opposite to a 
planar sensor. The small inter-electrode distance allows to reduce the charge carrier trapping probability after irradiation, 
acquire the induced charge faster, and use low voltages to fully deplete the sensor, among other benefits~\cite{3d-seminal}. 
This low depletion voltage enables to operate the 3D sensors with an extremely low power consumption, especially compared
with equivalent planar sensors.

However, the columnar structure also presents some disadvantages. The electric field is not uniform within the depleted
region, and therefore the sensor response cannot be uniform either. Indeed, there are charge losses inside the 
electrodes and ohmic columns, as the columns themselves are not active regions. Null points between same
type of electrodes will delay the created charges generated in there, as they have to diffuse first. 
As the inter-electrode distance is smaller, the capacitance of a 3D sensor is order of 3-5 
times higher than an equivalent planar, and consequently a 3D sensor is noisier. The fabrication process is still in the development phase,
being an expensive and time consuming process made by just a few R\&D 
institutions or foundries (IMB-CNM in Barcelona, FBK in Trento or SINTEF in Oslo), especially compared to the 
manufacturing of planar sensors. There are many companies capable to produce planar sensors, and 
with a well-established production, even for this small-pitch detector.

Several of these drawbacks can be or have been overcome. For instance, the production processes have
been simplified and improved over the past few years, in particular by using single-sided processes 
as opposed to the double-sided initially used for the first 3D batches~\cite{3d-doubledside}. The latest
productions are mature enough to achieve overall good yields~\cite{3d-singledside}. Another example:
the response of the electric field can be \emph{uniformed} by tilting the detector so that the particle 
passes through different electric field regions.

The 3D pixel technology has already been successfully used in some LHC experiments as position tracking detectors
in radiation environments such as in the ATLAS IBL~\cite{atlas-ibl}, the ATLAS Forward Proton (AFP) 
detector~\cite{atlas-afp}, and the CMS-TOTEM Precision Proton Spectrometer (PPS) detector~\cite{cms-totem-pps};
proving to be a mature enough yet novel technology. For the HL-LHC, ATLAS already decided to include 3D 
pixel detectors with \SI[product-units=power]{50 x 50}{\micro\meter} and 
\SI[product-units=power]{25 x 100}{\micro\meter} pixel size in the innermost layer of its new silicon inner 
tracker~\cite{ph2-atlas}. At the time of this report, CMS is in the process of deciding the inclusion 
of this technology in the two innermost layers of the barrel and in the inner ring of the end-cap of its
inner tracking detector.

The RD50 Collaboration has been involved in several activities related with the 3D pixel technology, such 
as their time performance in order to be used as timing detectors, the radiation hardness at extreme
fluences larger than \SI{1e17}{\neq\square\per\centi\meter}, and the qualification and characterization of the 
small-pitch 3D sensor with the pre-production readout chip RD53A~\cite{rd53a}, as part of the sensor 
qualification process for the ATLAS and CMS Phase-2 upgrade programs. The RD50 collaboration has also played 
an important role to bring 3D sensors to industrial production. In this paper, a summary of 
the last line of work is being reported. Information and results of the other work packages can be
found elsewhere~\cite{rd50-other,rd50-other-2,rd50-extreme-fluence}.

\section{Sensor characterization for the HL-LHC experiments upgrade}
\begin{figure}[tbh]
    \centering
    \includegraphics[width=0.5\textwidth]{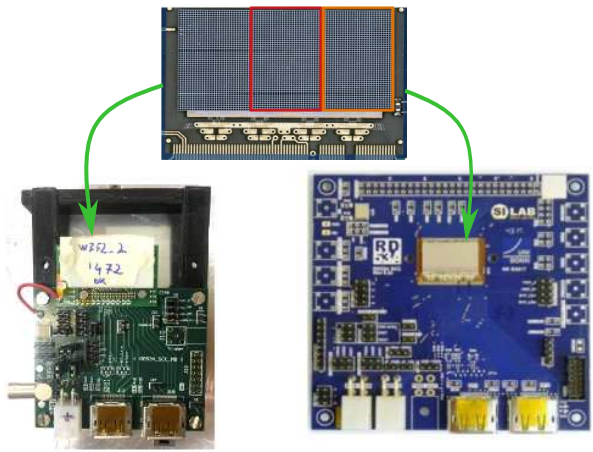}
    \caption{A RD53A chip picture (top) with the linear and differential regions highlighted in red and
             orange respectively. The chip bump-bonded to a sensor, is wire-bonded to a single chip card
             as part of the characterization setup. Two different single chip cards have been used, the 
             RICE card (left, developed by RICE University), and the Bonn card (right, developed by 
             the SiLab group at Bonn University).}
    \label{fig:rd53a}
\end{figure}

The inner tracker upgrade projects of ATLAS and CMS require small-pitch, radiation-hard detector
candidates, to be assembled with the readout electronics and mechanics projected by the collaborations. The 
readout chip (ROC) developed for CMS and ATLAS, the RD53A, is a demonstrator ROC for CMS and ATLAS upgrades, 
based on \SI{65}{\nano\meter} CMOS process~\cite{rd53a}. This is a common layout design developed by 
the RD53 collaboration, which included several Analogs Front-Ends (AFE) to be tested. The CMS and ATLAS
ROCs prototypes, the CROC and the ITkPixV1, respectively, have finally chosen to use different AFEs:
the linear AFE for CROC and the differential AFE for ITkPixV1. Fig.~\ref{fig:rd53a} shows a picture of the 
RD53A ROC with the linear and differential front-end regions highlighted, and the single chip cards assemblies
used for the characterization.

\subsection{Module description}
The sensors used  have been produced either by IMB-CNM in Barcelona (Spain) or by the 
FBK in Trento (Italy). In both cases, the production was made with a single-sided process on
p-type substrates Silicon on Silicon (SiSi) or Silicon on Insulator (SOI). The former has finally been 
chosen over the SOI, as it does not need the removal of the handle-wafer for biasing, which simplifies 
the production process and improves the yield. FBK used 6 inches Direct Bond Wafers
(DBW) with a support wafer made of \SI{500}{\micro\meter} thick Czochralski (CZ) silicon, with less than
\SI{1}{\ohm\centi\meter} resistivity. The electrodes were implanted by a Deep Reactive Ion Etching (DRIE)
process, in a Float Zone (FZ) wafer of high resistivity (higher than \SI{3000}{\ohm\centi\meter}) of 
\SI{130}{\micro\meter} (production with Mask Aligner lithography) and \SI{150}{\micro\meter} (production
with Stepper lithography) thickness. The columns have about \SI{5}{\micro\meter} of diameter.

For the case of the IMB-CNM, the p-type substrates were produced on 4 inches SOI wafers or 4 inches SiSi wafers 
with a handle wafer of \SI{300}{\micro\meter} thick CZ silicon of low resistivity. The electrodes are 
implanted on an active region with a nominal resistivity between 1000-\SI{5000}{\ohm\centi\meter} and 
\SI{150}{\micro\meter} thick.  On the SiSi wafers, there is no need to remove the handle wafer for bias, 
as the low-resistivity wafer is conductive. The columns have a diameter of about \SI{8}{\micro\meter}. 

\begin{figure}[tbh]
    \centering
    \begin{minipage}[c]{0.45\textwidth}
        \centering
        \includegraphics[width=1.0\textwidth]{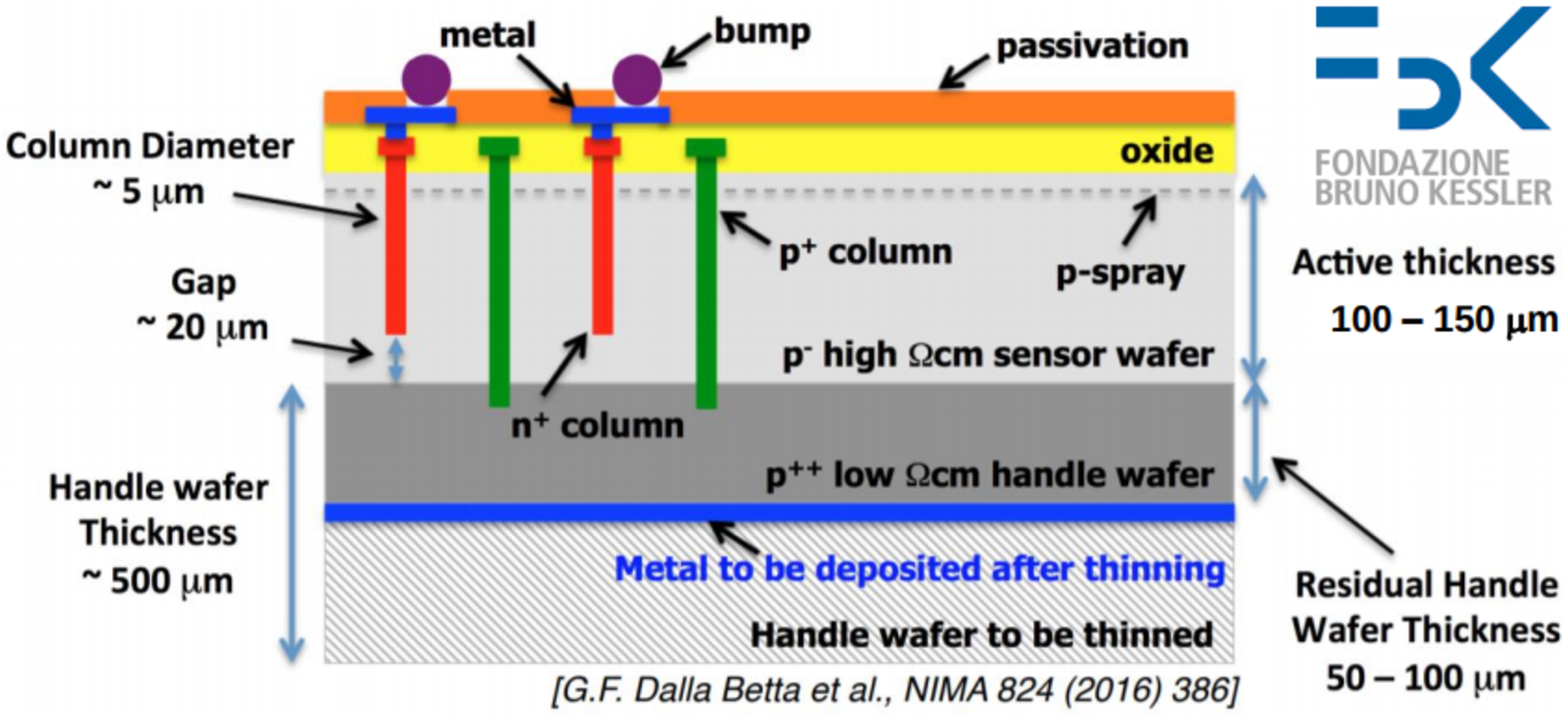} \\
        (a) FBK
    \end{minipage}
    \begin{minipage}[c]{0.45\textwidth}
        \centering
        \includegraphics[width=1.0\textwidth,height=3.3cm]{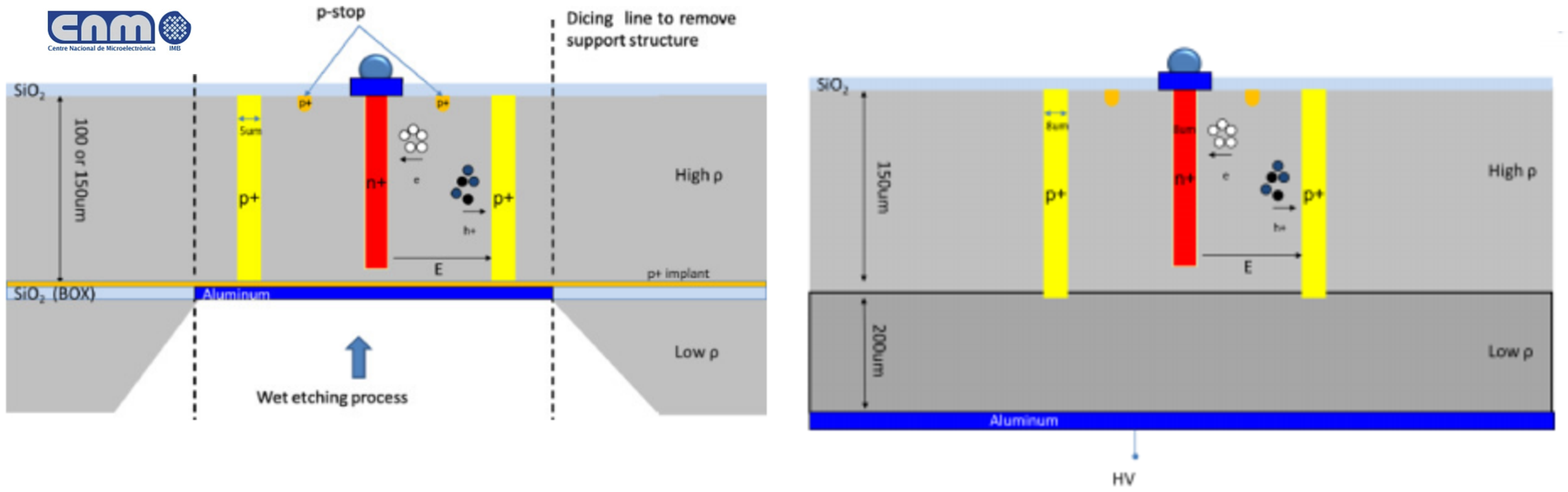} \\
        (b) IMB-CNM, SOI (left) and SiSi (right)
    \end{minipage}
    \caption{Sensor cross-section layout. The main components are schematized, showing the
             active region (light gray, high-resistive layer), the support CZ layer 
             (dark gray), the highly-doped column implant (red) and the ohmic columns (green,
             yellow).}
    \label{fig:sensor-layout}
\end{figure}

Fig.~\ref{fig:sensor-layout} summarizes schematically the described sensors.

\begin{figure}[tbh]
    \centering
    \includegraphics[width=0.6\textwidth]{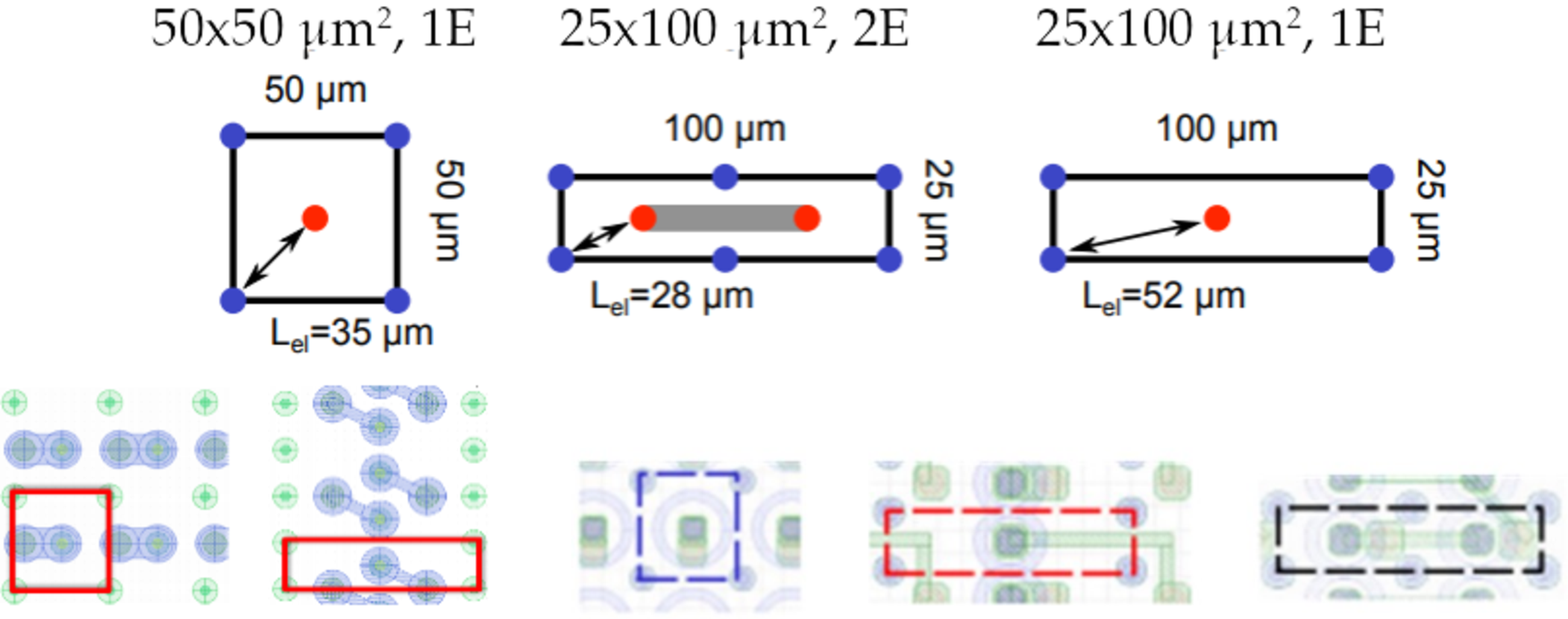}
        \caption{Pixel geometries considered for CMS and ATLAS. Each pixel cell is defined by the
        ohmic columns (blue) and the highly doped implant electrode (red). The inter-electrode
        distance is also specified. The bottom row is showing the different routing pattern used 
        to connect the bond-bumps on the sensor with the RD53A readout electrodes. Solid 
        (for the FBK modules) and dashed (for the IMB-CNM) lines are superimposed to stress
        a pixel cell.}
    \label{fig:pixel-geo}
\end{figure}

The sensors were manufactured with rectangular and square geometries, matching 
the layouts CMS and ATLAS are considering to use. The square \SI[product-units=power]{50 x 50}{\micro\meter}
has an inter-electrode distance of about \SI{35}{\micro\meter}, and is the easiest to produce.
Furthermore, they do not suffer from the cross-talk originated in the metallic routes~\cite{xtalk-25x100}, 
needed to match the RD53A \SI[product-units=power]{50 x 50}{\micro\meter} bond matrix pattern. 
However, a rectangular geometry allows to increase the resolution in the short pitch side, and therefore
were considered as well. Two rectangular geometries \SI[product-units=power]{25 x 100}{\micro\meter} have
been tested: with 2 electrodes (2E) per pixel cell with a very small inter-electrode distance 
of \SI{28}{\micro\meter} but difficult to produce, and 1 electrode (1E), less problematic to produce but
with larger inter-electrode distance. The later has been proven to be enough radiation hard, even with its
\SI{52}{\micro\meter} of inter-electrode distance, and therefore the 2E flavour has already been discarded.
Fig.~\ref{fig:pixel-geo} summarizes all the pixel geometries used in the results presented in this report.

Some sensors were also irradiated in order to emulate the damage produced by irradiation after years 
of operation in the experiments. Irradiations have been carried out at several facilities, with 
different particles and energies, in several intermediate steps until reaching fluences up to
\SI{1e16}{\neq\square\per\centi\meter}, the expected fluence of the innermost layers of CMS or ATLAS
after the two first runs of the HL-LHC in the scenario of \SI{3000}{\per\femto\barn}. Details about
the irradiations and methodology can be found in Ref.~\cite{irradiations}.

\subsection{Test beam characterization}
Electrical, TCT, TCT-TPA, or other characterizations techniques were used to study the sensors, 
and results have already been extensively presented elsewhere~\cite{cv-iv-1,hit-eff,rd50-extreme-fluence}. The sensors described 
in the previous section were bump-bonded to the RD53A demonstrator chip and wire bonded to PCB cards as explained 
before, and tested under a beam of particles. Two main facilities were used: CERN~\cite{cern-tb-facility}, 
with hadron beams of \SI[per-mode=symbol]{120}{\giga\electronvolt\per\clight}, and 
DESY~\cite{desy-tb-facility}, where electrons of \SI[per-mode=symbol]{5.6}{\giga\electronvolt\per\clight} 
are delivered to the beam lines. Test beam experiments are useful to study the behaviour and response
of a sensor under similar conditions it will face when it is mounted in the real experiment. The sensor
response is measured with the help of a telescope~\cite{aida-telescopes}, where the incident particles
are identified and reconstructed. Telescope reconstructed particles are used to identify the position
in the sensor under study where a hit should appear, allowing to measure hit-efficiency, charge, 
resolution, etc.

\begin{figure}[tbh]
    \centering
    \includegraphics[width=.7\textwidth]{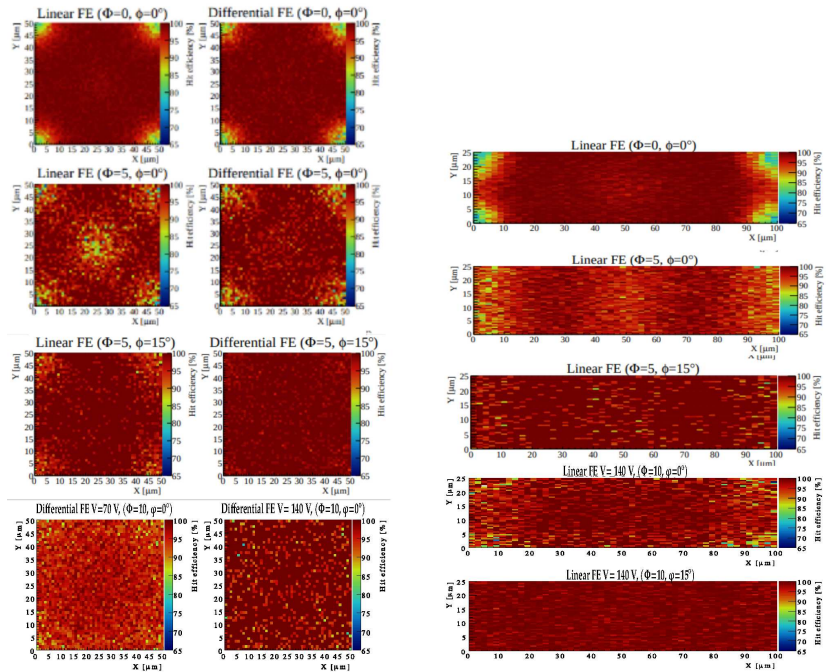}
        \caption{Hit efficiency per pixel cell in different sensors, shown from left to right; 50x50 Linear
        and Differential front-end, and 25x100 Linear front-end. From top to bottom are with increasing
        fluence ($\Phi$) and incident angle of the tracks ($\phi$). Details are inserted in the figures, 
        with fluence in units of \SI{1e15}{\neq\square\per\centi\meter} (0 for fresh sensors). 
        The operating bias voltaged used is also shown in some plots. Sensors were produced at IBM-CNM 
        with SOI p-type substrates. Figure re-adapted and composed from Ref.~\cite{hit-eff}.}
    \label{fig:hit-efficiency}
\end{figure}
The first important observable reported is the efficiency per pixel cell, measured for 
\SI[product-units=power]{25 x 100}{\micro\meter} and \SI[product-units=power]{50 x 50}{\micro\meter} fresh sensors. 
The two plots in Fig.~\ref{fig:hit-efficiency} show the hit efficiency drop in the ohmic columns 
at the four corners when the incident particles enter perpendicular to the sensor
plane. This drop is the result of the incident particle passing inside the non-active region of the 
columns. However, there is no significant drop observed in the n-column at center, as the charge 
generated below the column (see Fig.~\ref{fig:sensor-layout}) is enough to create a hit 
(thresholds are tuned around 1000 electrons).

\begin{figure}[tbh]
    \centering
    \begin{minipage}[c]{0.45\textwidth}
        \centering
        \includegraphics[width=.7\textwidth]{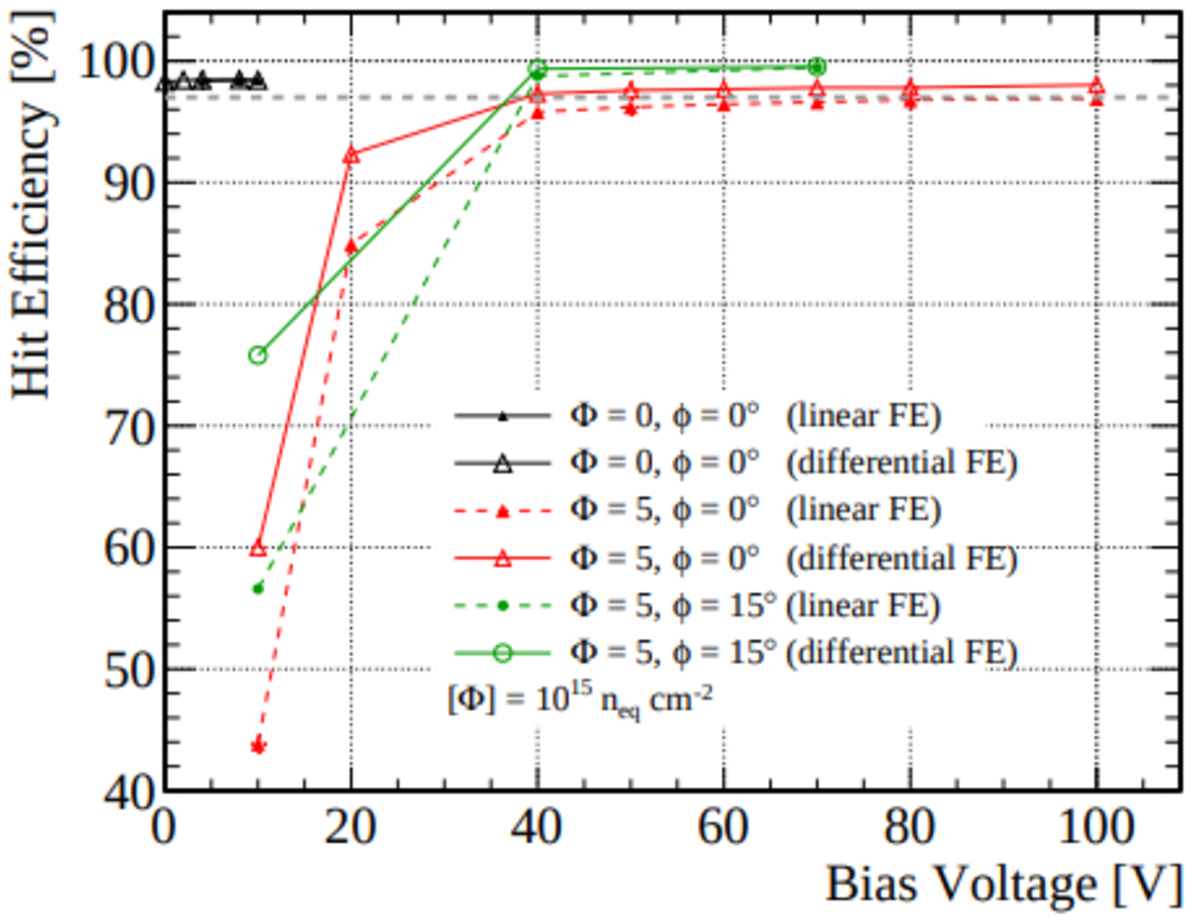} \\
        (a) 50x50, fresh and \SI{5e15}{\neq\square\per\centi\meter }
    \end{minipage} 
    \begin{minipage}[c]{0.45\textwidth}
        \centering
        \includegraphics[width=.7\textwidth]{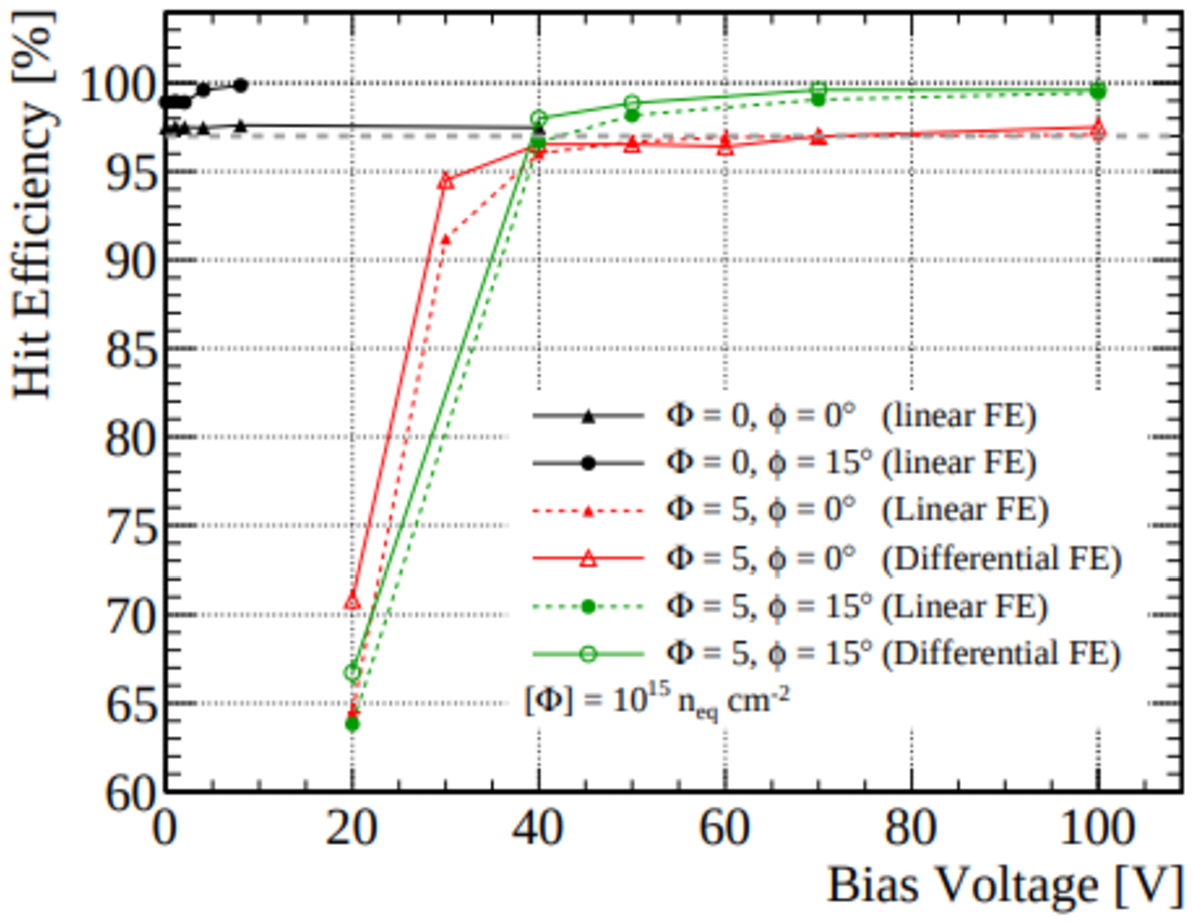} \\
        (b) 25x100, fresh and \SI{5e15}{\neq\square\per\centi\meter}
    \end{minipage} \\
    \begin{minipage}[c]{0.45\textwidth}
        \centering
        \includegraphics[width=0.7\textwidth]{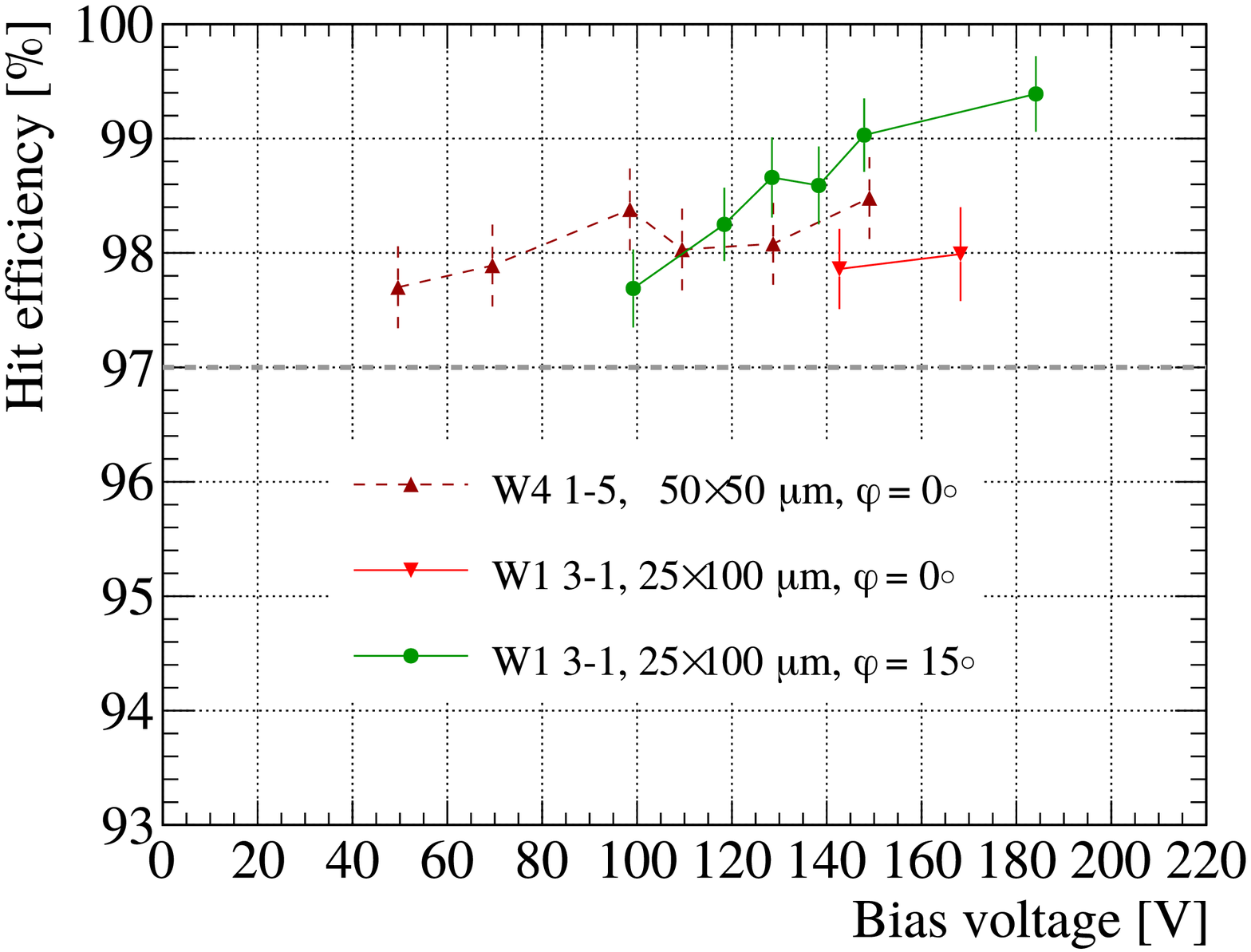} \\
        (c) 50x50 and 25x100,  \SI{1e16}{\neq\square\per\centi\meter}
    \end{minipage}
    %\begin{minipage}[c]{0.45\textwidth}
    %    \centering
    %    \includegraphics[width=.6\textwidth]{eff_volt_50by50_1E16.eps} \\
    %    (c) 50 x 50, \SI{1e16}{\neq\square\per\centi\meter}
    %\end{minipage}  
    %\begin{minipage}[c]{0.45\textwidth}
    %    \centering
    %    \includegraphics[width=.7\textwidth,height=3.1cm]{eff_volt_25by100_1E16.eps} \\
    %    (d) 25 x 100, \SI{1e16}{\neq\square\per\centi\meter}
    %\end{minipage}
    \caption{Measured overall hit efficiency with respect the applied bias voltage, for fresh 
    and irradiated sensors. Figures extracted from~\cite{hit-eff}.}
    \label{fig:bias-voltage}
\end{figure}

The drop in the n-column appears in the irradiated sensor as consequence of two effects: thresholds are 
increased to deal with a noisier detector, and trapping, lowering the amount of generated charge 
below the column. Fresh sensors are exhibiting a 98\% overall efficiency, including the ohmic column losses. 
When tilted, a 99.5\% overall efficiency is recovered with very low bias voltages. As it can be observed
from Fig.~\ref{fig:bias-voltage}, fully depletion is reached with a few volts for fresh sensors, and with a few 
hundred of volts for sensors irradiated up to \SI{1e16}{\neq\square\per\centi\meter}.

\begin{figure}[tbh]
    \centering
    \begin{minipage}[c]{0.45\textwidth}
        \centering
        \includegraphics[width=.5\textwidth]{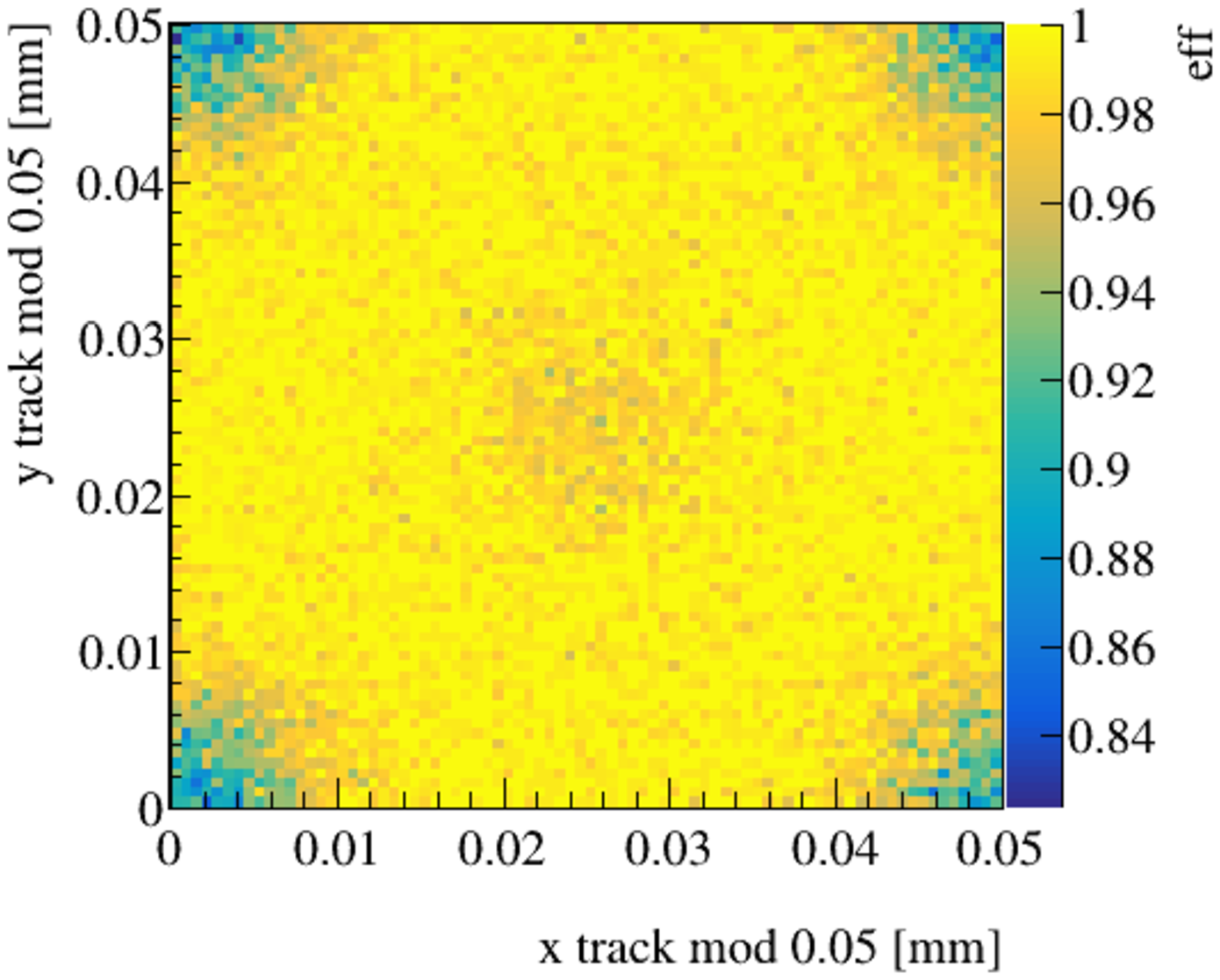} \\
    \end{minipage} 
    \begin{minipage}[c]{0.45\textwidth}
        \centering
        \includegraphics[width=.9\textwidth]{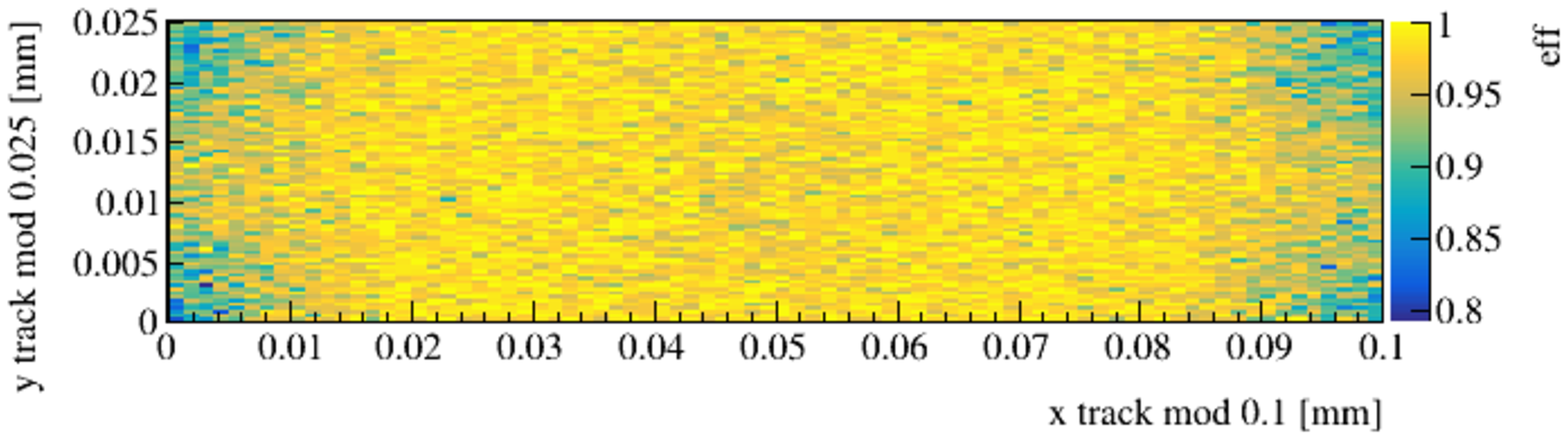} \\
    \end{minipage} \\
    \begin{minipage}[c]{0.45\textwidth}
        \centering
        \includegraphics[width=0.5\textwidth]{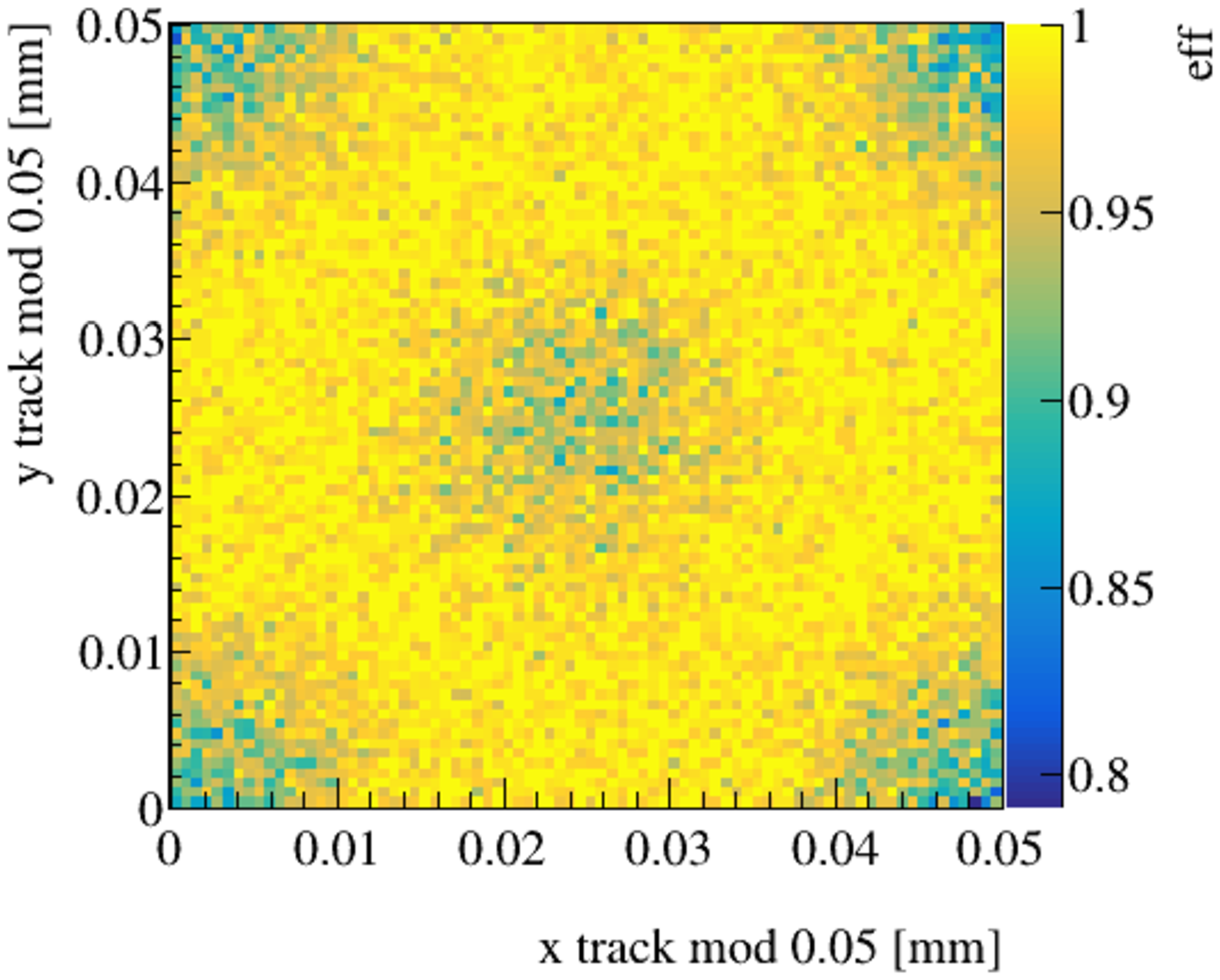} \\
    \end{minipage}  
    \begin{minipage}[c]{0.45\textwidth}
        \centering
        \includegraphics[width=.9\textwidth]{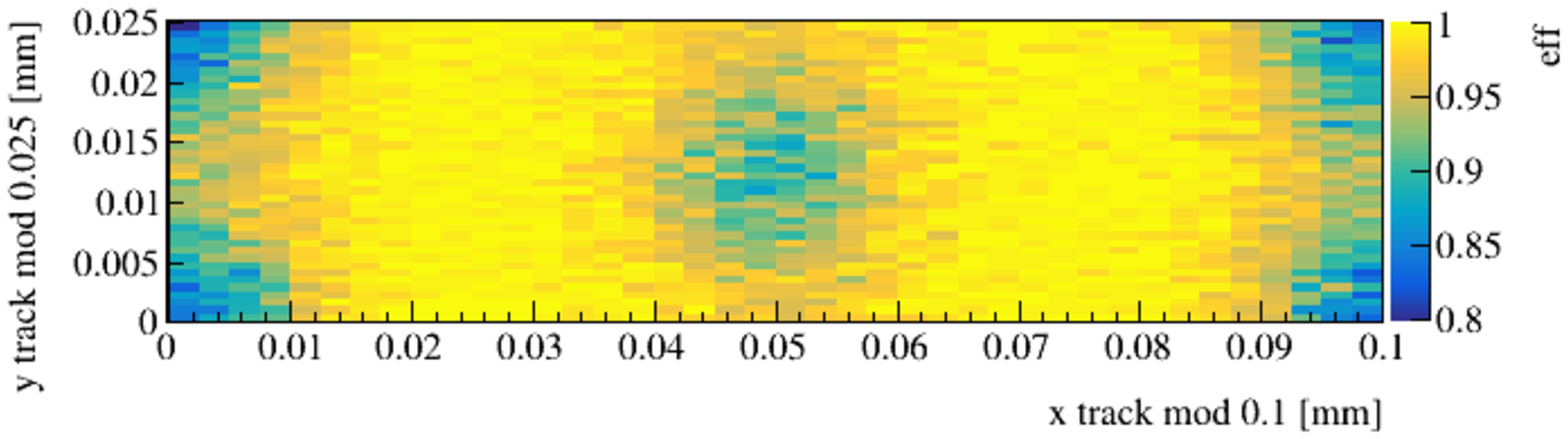} \\
    \end{minipage}
    \caption{Hit efficiency in pixel cell for fresh \SI[product-units=power]{50 x 50}{\micro\meter} (top-left)
            \SI[product-units=power]{25 x 100}{\micro\meter} (top-right); and after
            irradiation up to \SI{1e16}{\neq\square\per\centi\meter} for \SI[product-units=power]{50 x 50}{\micro\meter}
            (bottom-left) and \SI[product-units=power]{25 x 100}{\micro\meter} (bottom-right). Sensors produced
            by FBK.}
    \label{fig:hit-efficiency-fbk}
\end{figure}

Equivalent results have been obtained for FBK produced sensors~\cite{hit-eff-fbk}, showing similar 
performances (see Fig.~\ref{fig:hit-efficiency-fbk}), as well as similar depletion voltages: 
sensors irradiated at \SI{1e16}{\neq\square\per\centi\meter} fully depletes around 
\SIrange[range-phrase=--,range-units=single]{120}{150}{\volt}.  

The resolution of the sensor attached to the RD53A before and after irradiation up to 
\SI{1e16}{\neq\square\per\centi\meter}, has been recently measured in the short pixel direction of the
\SI[product-units=power]{25 x 100}{\micro\meter} geometry~\cite{resolution}.  The sensor shows a resolution of
\SI{3}{\micro\meter} and there is almost no degradation after irradiation, with 
a resolution of \SI{5.7}{\micro\meter} evaluated at the optimal angle. These results complement
the previously obtained for the \SI[product-units=power]{50 x 50}{\micro\meter} geometry, with 
a resolution of \SI{5}{\micro\meter}, evaluated at the optimal angle and before irradiation.

\begin{figure}[tbh]
    \centering
    \includegraphics[width=.7\textwidth]{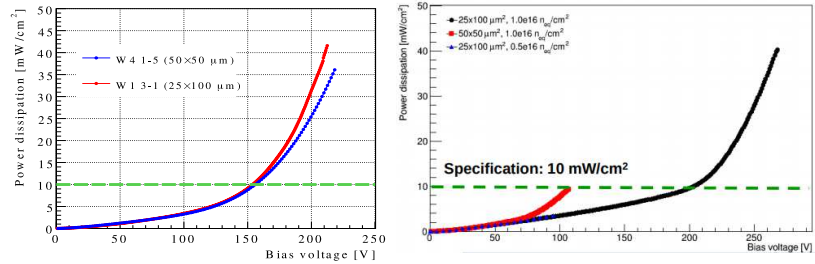}
    \caption{Sensor power dissipation with respect to the operating bias voltage for 
             25x100 and 50x50 geometries, irradiated up to \SI{1e16}{\neq\square\per\centi\meter}
             and operated with a temperature around \SI{-25}{\celsius}. Figure on the left corresponds
             to SOI IMB-CNM of \SI{150}{\micro\meter} thick sensors; on the right, FBK SiSi
              \SI{150}{\micro\meter} thick using the Stepper lithography mask.}
    \label{fig:power-comsumption}
\end{figure}

A parameter of interest, especially important for the commissioning and operation of the 
modules in the targeted experiment, is the power consumption. The ATLAS specifications
are requiring to be below \SI[per-mode=symbol]{10}{\milli\watt\per\square\centi\meter}, 
while CMS requires less than \SI[per-mode=symbol]{1}{\watt\per\square\centi\meter} (included sensor and chip consumption). The power
consumption of the tested sensors after being irradiated up to \SI{1e16}{\neq\square\per\centi\meter}, 
has been measured below the ATLAS specifications in the full operational range of the sensor, defined
with an overall efficiency larger than 97\%. Consumption has been evaluated operating the sensors at around 
\SI{-25}{\celsius}, and for both geometries \SI[product-units=power]{50 x 50}{\micro\meter}
and \SI[product-units=power]{25 x 100}{\micro\meter}. Fig.~\ref{fig:power-comsumption} summarizes
the results detailed in~\cite{hit-eff,pw-comp-1}.

\section{Conclusions}
Small-pitch 3D pixel sensors bump-bonded to the pre-production ROC RD53A have proven to have 
an excellent radiation tolerance up to \SI{1e16}{\neq\square\per\centi\meter}, keeping hit efficiency
larger than 97\% at particle normal incidence with operational bias voltages below \SI{200}{\volt}.
This allows to fulfill the power dissipation specifications required by the ATLAS and CMS detectors.

Both geometries being considered,  \SI[product-units=power]{25 x 100}{\micro\meter} and 
\SI[product-units=power]{50 x 50}{\micro\meter}, show similar performance, in particular 
the \SI[product-units=power]{25 x 100}{\micro\meter} with 1 electrode is able to deal with 
the radiation tolerance demanding conditions, therefore there is no need for the 2 electrodes design. 
Their performance meets the baseline requirements to be qualified for the HL-LHC inner tracker 
upgrades of ATLAS and CMS. 

A new set of test beam measurements is being planned for samples already 
irradiated to \SI{2e16}{\neq\square\per\centi\meter}, while further irradiation are planned to 
reach the expected fluences for the LS5, i.e. \SI{2.6e16}{\neq\square\per\centi\meter}, when is 
planned the replacement of the modules of the innermost layers of CMS. New productions of sensors
bump-bonded to the experiment-dedicated ROCs, the ItkPixV1 (ATLAS) and the CROC (CMS) have
already started.


\begin{thebibliography}{25}
    \bibitem{hl-lhc} G. Apollinari et al., \emph{High-Luminosity Large Hadron Collider (HL-LHC): Preliminary Design Report},
        CERN Yellow Reports: Monographs, doi:10.23731/CYRM-2017-004 (2017).
    \bibitem{ph2-cms} The CMS Collaboration, \emph{The Phase-2 Upgrade of the CMS Tracker - Technical Design Report}, 
        CERN-LHCC-2017-009, CMS-TDR-014 (2017).
    \bibitem{ph2-atlas} The ATLAS Collaboration, \emph{Technical Design Report for the ATLAS Inner Tracker Pixel Detector},
        CERN-LHCC-2017-021, ATLAS-TDR-030 (2017).
    \bibitem{3d-seminal} S. J. Parker et al., \emph{3D- A proposed new architecture for solid-state radiation detectors},
        Nucl. Instrum. Meth. A \textbf{395} 328 (1997).
    \bibitem{3d-doubledside} G. Pellegrini et al., \emph{3D double sided detector fabrication at IMB-CNM}, 
            Nucl. Instrum. Meth. A \textbf{699} (2013) 27. 
    \bibitem{3d-singledside} G. Pellegrini et al., \emph{3D-Si single sided sensors for the innermost layer of the ATLAS pixel upgrade},
        Nucl. Instrum. Meth. A \textbf{924} (2019) 69-72.
    \bibitem{atlas-ibl} ATLAS IBL Collaboration, \emph{Prototype ATLAS IBL Modules using the FE-I4A Front-End Readout Chip}, 
        JINST \textbf{7} (2012) P11010.
    \bibitem{atlas-afp} ATLAS Collaboration, \emph{Technical Design Report for the ATLAS Forward Proton Detector}, 
        CERN-LHCC-2015-009, ATLAS-TDR-024,  (2015).
    \bibitem{cms-totem-pps} CMS and TOTEM Collaborations, \emph{CMS-TOTEM Precision Proton Spectrometer}, 
        CERN-LHCC-2014-021, TOTEM-TDR-003, CMS-TDR-13 (2014).
    \bibitem{rd53a} M. Garcia-Sciveres, \emph{The RD53A Integrated Circuit}, CERN-RD53-PUB-17-001 (2017).
    \bibitem{rd50-other} M. Moll, \emph{RD50 Status Report}, 139th LHCC Meeting, {http://cern.ch/go/SQl7}, Accessed: 2020-11-16.
    \bibitem{rd50-other-2} G. Casse and M. Moll, \emph{RD50 Prolongation Request 2018}, 
        CERN-LHCC-2018-017. LHCC-SR-007 (2018).
    \bibitem{rd50-extreme-fluence} M. Manna et al., \emph{First characterisation of 3D pixel detectors irradiated at extreme fluences},
        Nucl. Instrum. Meth. A \textbf{979} (2020) 164458.
    \bibitem{xtalk-25x100} A. Garcia Alonso et al., \emph{Test beam characterization of irradiated 3D pixel sensors},
        Contribution to 21st iWoRiD {https://indico.cern.ch/event/774201/contributions/3429263/} (2019).
    \bibitem{irradiations} J. Duarte-Campderros, \emph{Study of 3D pixel sensors after non-uniform proton irradiation},
        Contribution to TREDI2020 {https://indico.cern.ch/event/813597/contributions/3727824/} (2020).
    \bibitem{cv-iv-1} I. Vila Alvarez et al., \emph{Update on CSIC activities on LGAD \& 3D sensors}, 
        Contribution to 4th AIDA-2020 Anual Meeting {http://cern.ch/go/7RGX} (2019).
    \bibitem{cern-tb-facility} \emph{AD, PS and SPS Users Schedules}, Web page {https://ps-sps-coordination.web.cern.ch/ps-sps-coordination/pindex.html}
        Accessed 2020-11-18.
    \bibitem{desy-tb-facility} R. Diener et al., \emph{The DESY II test beam facility}, 
        Nucl. Instrum. Meth. A \textbf{922} (2019) 265-288.
    \bibitem{aida-telescopes} I. Rubinskiy, on behalf of EUDET, AID Aconsortia, \emph{An EUDET/AIDA Pixel Beam Telescope for Detector Development},
        Phys. Procedia \textbf{37} (2012) 923-931.
    \bibitem{hit-eff} S. Terzo et al., \emph{A new generation of radiation hard 3D pixel sensors for the ATLAS upgrade},
        Nucl. Instrum. Meth. A \textbf{982} (2020) 164587.
    \bibitem{hit-eff-fbk} J. Duarte-Campderros et al., \emph{Results on proton-irradiated 3D pixel sensors interconnected to RD53A readout ASIC},
        Nucl. Instrum. Meth. A \textbf{944} (2019) 162625. 
    \bibitem{resolution} M. Meschini et al., \emph{Radiation resistant innovative 3D pixel sensors for the CMS upgrade at the High Luminosity LHC},
        Nucl. Instrum. Meth. A \textbf{978} (2020) 164429.
    \bibitem{pw-comp-1} A. A. Samy et al., \emph{Characterization of FBK 3D Pixel Sensor Modules Based on RD53A Readout Chip for the ATLAS ITk},
        Contribution to TREDI2020, {https://indico.cern.ch/event/813597/contributions/3727841} (2020).
\end{thebibliography}
\end{document}